\documentclass{aastex}
\usepackage{emulateapj5,epsfig}

\newcommand{\be}{\begin{equation}}
\newcommand{\ee}{\end{equation}}
\newcommand{\lp}{\left(}
\newcommand{\rp}{\right)}
\newcommand{\lb}{\left[}
\newcommand{\rb}{\right]}
\newcommand{\etal}{et al.}

\newcommand{\mpr}{m_{\rm p}}
\newcommand{\me}{m_{\rm e}}
\newcommand{\Ye}{Y_{\rm e}}

\newcommand{\Batom}{B_{\rm 0}}
\newcommand{\Bq}{B_{\rm Q}}

\newcommand{\bqnperp}{s}
\newcommand{\bqnz}{\nu}
\newcommand{\sigth}{\sigma_{\rm T}}
\newcommand{\omegab}{\omega_{Be}}
\newcommand{\omegabp}{\omega_{Bp}}

\newcommand{\omegap}{\omega_{{\rm p}e}}
\newcommand{\vel}{v_{\rm e}}
\newcommand{\Ebp}{E_{Bp}}
\newcommand{\Ebi}{E_{Bi}}

\newcommand{\Evp}{E_{\rm V}}
\newcommand{\densvp}{\rho_{\rm V}}
\newcommand{\kabsj}{\kappa^{\rm abs}_j}
\newcommand{\kabsa}{\kappa^{\rm abs}_\alpha}
\newcommand{\kabsx}{\kappa^{\rm abs}_X}
\newcommand{\keso}{\kappa^{\rm es}_0}
\newcommand{\polarevectwo}{|e_\alpha^j|^2}
\newcommand{\polarevecthree}{{\bf e}^j}
\newcommand{\veck}{\hat k}
\newcommand{\vecB}{\hat B}

\newcommand{\Teff}{T_{\rm eff}}
\newcommand{\Teffsix}{T_{\rm eff,6}}
\newcommand{\Kabsx}{K^{\rm abs}_X}
\newcommand{\Kabsj}{K^{\rm abs}_j}
\newcommand{\Kabsp}{K^{\rm abs}_+}
\newcommand{\Kabsm}{K^{\rm abs}_-}
\newcommand{\Ena}{E_{n-1}}
\newcommand{\Enb}{E_{n+1}}
\newcommand{\Enc}{E_{n}}
\newcommand{\holaione}{HL01}
\newcommand{\holaitwo}{HL03}
\newcommand{\laiho}{LH02}
\newcommand{\potcha}{PC}

\shorttitle{Partially ionized NS atmospheres}
\shortauthors{W.C.G. Ho, D. Lai, A.Y. Potekhin, and G. Chabrier}

\begin{document}

\title{Atmospheres and spectra of strongly magnetized neutron stars -- III.
 Partially ionized hydrogen models}
\author{Wynn C. G. Ho and Dong Lai}
\affil{Center for Radiophysics and Space Research, 
Department of Astronomy, Cornell University,
Ithaca, NY 14853}
\email{wynnho, dong@astro.cornell.edu}
\author{Alexander Y. Potekhin}
\affil{Ioffe Physico-Technical Institute,
Politekhnicheskaya 26, 194021, St. Petersburg, Russia;}
\affil{Isaac Newton Institute of Chile, St. Petersburg Branch, Russia}
\email{palex@astro.ioffe.rssi.ru}
\and
\author{Gilles Chabrier}
\affil{CRAL (UMR CNRS No. 5574), Ecole Normale Sup\'{e}rieure de Lyon,
69364 Lyon Cedex 07, France}
\email{chabrier@ens-lyon.fr}

%%%%%%%%%%%%%%%%%%%%%%%%%%%%%%%%%%%%%%%%%%%%%%%%%%%%%%%%%
\begin{abstract}
We construct partially ionized hydrogen atmosphere models for magnetized
neutron stars in
radiative equilibrium with surface fields $B=10^{12}-5\times 10^{14}$~G
and effective temperatures $\Teff\sim\mbox{a few}\times 10^5-10^6$~K.
These models are based on the latest equation of state and opacity
results for magnetized, partially ionized hydrogen plasmas that
take into account various magnetic and dense medium effects.
The atmospheres directly determine the characteristics of thermal
emission from isolated neutron stars.
For the models with $B=10^{12}-10^{13}$~G, the spectral features due to
neutral atoms lie at extreme UV and very soft X-ray energy bands and
therefore are difficult to observe.  However, the continuum
flux is also different from the fully ionized case, especially at
lower energies.
For the superstrong field models ($B\ga 10^{14}$~G), we show that
the vacuum polarization effect not only suppresses the proton
cyclotron line as shown previously, but also suppresses
spectral features due to bound species;
therefore spectral lines or features in thermal radiation are more
difficult to observe when the neutron star magnetic field is
$\ga 10^{14}$~G.
\end{abstract}


%%%%%%%%%%%%%%%%%%%%%%%%%%%%%%%%%%%%%%%%%%%%%%%%%%%%%%%%%%%%%%%%%%%
\section{Introduction} \label{sec:intro}

Thermal radiation from the surface of isolated neutron stars (NSs)
can provide invaluable information on the physical properties and
evolution of NSs.  In the last few years, such radiation has been
detected from radio pulsars (see Becker~2000;
Pavlov, Zavlin, \& Sanwal~2002), from old and young radio-quiet NSs
(see Treves \etal~2000; Pavlov \etal~2002), and from
soft gamma-ray repeaters (SGRs) and anomalous X-ray pulsars (AXPs)
(see Hurley~2000; Mereghetti \etal~2002a),
which form a potentially new class of NSs
(``magnetars'') endowed with superstrong ($B\ga 10^{14}$~G) magnetic
fields (see Thompson \& Duncan~1996; Thompson~2001).
The NS surface emission is
mediated by the thin atmospheric layer (with scale height
$\sim 0.1-10$~cm and density $\sim 0.1-10^3\mbox{ g cm$^{-3}$}$) that
covers the stellar surface.  Therefore, to properly interpret
the observations of NS surface emission and to provide
accurate constraints on the physical properties of NSs, it is
important to understand in detail the radiative properties
of NS atmospheres in the presence of strong magnetic fields
(see Pavlov \etal~1995; Ho \& Lai~2001, 2003a; Zavlin \& Pavlov~2002
for more detailed references on observations and on previous works
of NS atmosphere modeling).

This paper is the third in a series where we systematically
investigate the atmosphere and spectra of strongly magnetized NSs.
In Ho \& Lai~(2001, hereafter \holaione), we constructed
self-consistent NS atmosphere models in radiative equilibrium
with magnetic field $B\sim 10^{12}-10^{15}$~G and effective
temperature $\Teff\sim 10^6-10^7$~K and assuming the atmosphere
is composed of fully ionized hydrogen or helium;
we focused on the effect of the ion cyclotron resonance at
$E=\Ebi=0.63\,\Ye(B/10^{14}\mbox{ G})$~keV, where
$\Ye=Z/A$ is the electron fraction and $Z$ and $A$
are the atomic charge and atomic mass of the ion, respectively,
and showed that the spectra can exhibit a broad feature due
to the ion resonance (see also Zane \etal~2001).
In Ho \& Lai~(2003a, hereafter \holaitwo; see also Lai \& Ho 2002, 2003),
we studied the effect
of vacuum polarization on the atmosphere structure and radiation
spectra of strongly magnetized NSs.  Polarization of the vacuum
due to virtual $e^+e^-$ pairs becomes significant
when $B\ga\Bq$, where $\Bq=\me^2c^3/e\hbar=4.414\times 10^{13}$~G
is the critical QED magnetic field strength.
Vacuum polarization modifies the dielectric property of the medium
and the polarization of photon modes (e.g., Adler~1971;
Tsai \& Erber~1975; Gnedin, Pavlov, \& Shibanov~1978;
Heyl \& Hernquist~1997), thereby altering the radiative scattering
and absorption opacities (e.g., M\'{e}sz\'{a}ros \& Ventura~1979;
Pavlov \& Gnedin~1984; see M\'{e}sz\'{a}ros~1992 for review).
We showed that
vacuum polarization leads to a broad depression in the high-energy 
($E\ga$ a few keV) radiation flux from the atmospheres and hence
softer high energy tails, as compared
to models without vacuum polarization.  The depression in the
spectrum is broad because of the large density gradient in the atmosphere.
This depression of continuum flux also strongly suppresses
the equivalent width of the ion cyclotron line and makes the line
more difficult to observe (for $B\ga\Bq$).
\holaitwo\ suggests that the absence
of lines in the observed thermal spectra of several AXPs may be an
indication of the vacuum polarization effect at work in these
systems.

In our previous studies, as well as in other previous studies of
magnetized hydrogen and helium NS atmospheres (see Pavlov \etal~1995;
Zavlin \& Pavlov~2002),
the atmospheres are assumed to be completely ionized.
Because the strong magnetic field significantly increases
the binding energies of atoms, molecules, and other bound states
(see Lai~2001 for a review), these bound states may have abundances
appreciable enough to contribute to the opacity in the atmosphere
(see Lai \& Salpeter~1997; Potekhin, Chabrier, \& Shibanov~1999).
Although the energy levels and radiative transitions of a stationary
H atom in strong magnetic fields have long been understood (e.g.,
Ruder \etal~1994), complications arise from the non-trivial
effect of the center-of-mass motion of the atom on its internal
structure (e.g., Potekhin~1998, and references therein).
Also, because of the relatively high atmosphere density
($\rho\ga 1\mbox{ g cm$^{-3}$}$), a proper treatment of pressure
ionization is important (e.g., Lai \& Salpeter~1995; Potekhin \etal~1999).
Recently, thermodynamically consistent equation of state
(EOS) and opacities for a magnetized, partially ionized H plasma
have been obtained by Potekhin \& Chabrier (2003, hereafter \potcha).
In this paper, we study NS atmosphere models based on the
EOS and opacities of \potcha\
(see Ho \etal~2003 for an initial report of the results shown here).

Section~\ref{sec:magatom} briefly summarizes the energy structure
of hydrogen atoms in strong magnetic fields.
Some details of the atmosphere models are discussed in
Section~\ref{sec:nummeth}.  We present atmosphere models and
spectra for different magnetic field strengths and temperatures
in Section~\ref{sec:results}.
In Section~\ref{sec:plasma}, we examine a dense plasma effect not
accounted for in previous modeling of NS atmospheres.
Section~\ref{sec:discussion} summarizes and discusses the
implications of our results.

%%%%%%%%%%%%%%%%%%%%%%%%%%%%%%%%%%%%%%%%%%%%%%%%%%%%%%%%%%%%%%%%%%%
\section{Hydrogen Atom in Strong Magnetic Fields} \label{sec:magatom}

In a strong magnetic field with $b=B/\Batom\gg 1$, where
$\Batom=\me^2e^3c/\hbar^3=2.3505\times 10^9$~G,
the magnetic field confines the electron gyromotion in the
transverse direction to a size $\sim a_0/\ln b$, where $a_0$
is the Bohr radius, and
the electron is restricted to the ground Landau level.
The hydrogen atom energy spectrum is then specified by two
quantum numbers ($\bqnperp,\bqnz$), where $\bqnperp$ measures the
mean transverse separation between the electron and proton
(or the guiding center distance of the electron gyromotion)
and $\bqnz$ gives the number of nodes in the $z$ wave function.
For $\ln[b/(2\bqnperp+1)] \gg 1$, the energy of the states with $\bqnz=0$
is $E_\bqnperp^0 = - {\cal A}_\bqnperp (\ln b)^2$, where the superscript
``0'' indicates the energy of an atom with a non-moving
(infinitely massive) nucleus and ${\cal A}_\bqnperp$ depends weakly on
$b$ and $\bqnperp$ and formally tends to 1~Ry ($=13.6$~eV) at $b\to\infty$.
For example, for $\bqnperp=0$ and
$10^{11}\mbox{ G}< B < 10^{16}\mbox{ G}$, ${\cal A}_\bqnperp = (5.0\pm0.7)$~eV.
A convenient fitting formula for {\em arbitrary} $B$ (accurate to within
a few parts in $10^3$ for $B<10^{17}$~G and $\bqnperp\leq 3$) is
\be
E_\bqnperp^0 = - {\cal A}_{\bqnperp} x^2, \qquad x = \ln(1+c_0 b),
\label{eq:enbh1}
\ee
\be
\frac{{\cal A}_{\bqnperp}}{\mbox{Ry}} = \frac{1+c_2x^{-2}+c_3 x^{-3}
 +c_5x^{-5}+c_6(1+s)^{-2} x^{-6}}{1+c_1x^{-1}+c_4x^{-2}+c_6 x^{-4} }.
\label{eq:enbh2}
\ee
Here, $c_5 = c_6(1+\bqnperp)/c_0$ ensures the correct behavior of the fit
at $b\ll 1$, and the parameters $c_i$ ($i=0,1,2,3,4,6$) are given in
Table~\ref{tab:enfit} for $\bqnperp=0,1,2,3$.

For the states with $\bqnz > 0$, the energies at $b\gg 1$ are
$E_{\bqnperp\bqnz}^0 \simeq n^{-2}$~Ry, where $n$ is the integer part
of $(\bqnz+1)/2$ (see Lai 2001, and references therein; see also
Potekhin 1998 for a more comprehensive set of fitting formulae).

\begin{table*}
\caption{Parameters for energy levels of hydrogen atom
[see eqs.~(\ref{eq:enbh1}) and (\ref{eq:enbh2})]
\label{tab:enfit}}
\begin{tabular}{ccccccc}
 $\bqnperp$ & $c_0$ & $c_1$ & $c_2$ & $c_3$ & $c_4$ & $c_6$ \\
 0 & 1 & 20.62 & 27.37 & 164.19 & 106.0 & 1175.6 \\
 1 & 5 & 19.57 & 19.72 & 212.0 & 495.5 & 1903.7 \\
 2 & 12 & 12.53 & 13.02 & 207.8 & 907.9 & 1532.6 \\
 3 & 21 & 3.393 & 8.177 & 203.6 & 1313.8 & 1140.1
\end{tabular}
\end{table*}

This simple picture of the H energy levels is modified when a finite
proton mass is taken into account.  Even for a ``stationary'' H atom,
the energy $E_\bqnperp$ becomes $E_\bqnperp^0+\bqnperp\hbar\omegabp$,
where $\hbar\omegabp=6.3\,B_{12}$~eV is the proton cyclotron energy
and $B_{12}=B/(\mbox{$10^{12}$ G})$;
the extra ``proton recoil'' energy $\bqnperp\hbar\omegabp$ becomes
increasingly important with increasing $B$.
Moreover, the effect of center-of-mass motion is non-trivial: When
the atom moves perpendicular to the magnetic field, a strong electric
field is induced in its rest frame and can significantly change the
atomic structure.  Indeed, in the limit of large transverse
pseudomomentum $p_\perp$, the atom assumes a decentered configuration,
and its energy depends on $p_\perp$ as $-p_\perp^{-1}$ rather than
the usual $p_\perp^2$ dependence
(see Potekhin~1998 for fitting formulae and references).

The most important transitions for the radiative opacity are the
bound-free transition from the $(\bqnperp,\bqnz)=(0,0)$ state and the
bound-bound transition $(0,0)\rightarrow(1,0)$.
The former produces an absorption edge at the energy $E_0^0$,
which is smoothed out due to the dependence of $E_0$ on $p_\perp$
from magnetic broadening (see Potekhin \& Pavlov~1997) and due to
the merging with transitions to $\bqnz>0$ states at slightly
lower energies (\potcha).
The $(0,0)\rightarrow(1,0)$ transition produces the maximum absorption
at the energy $E_0(p_\perp=0)-E_1(p_\perp=0)=E_0^0-E_1^0+\hbar\omegabp$;
with increasing $p_\perp$, $E_0(p_\perp)-E_1(p_\perp)$ first decreases
to a minimum and then increases to $\hbar\omegabp$.
As a result, this bound-bound transition produces a magnetically
broadened feature that resembles an inverted absorption edge, rather
than a line, plus peaks at $\min(E_0-E_1)$ and $\hbar\omegabp$
(Pavlov \& Potekhin~1995).
The transitions from $\bqnperp=0$ to $\bqnperp>1$
are forbidden for stationary atoms but become allowed for moving
atoms and thus contribute to the opacity.  See \potcha\ for more
detailed discussion of opacity features.

%%%%%%%%%%%%%%%%%%%%%%%%%%%%%%%%%%%%%%%%%%%%%%%%%%%%%%%%%%%%%%%%%%%
\section{Model Method} \label{sec:nummeth}

All models considered in this paper have the magnetic field aligned
perpendicular to the stellar surface, and thus the spectra presented
are emission from a local patch of the NS surface.
We solve the full, angle-dependent radiative transfer equations
for the two coupled photon modes in order to construct self-consistent
NS atmosphere models.
The two photon polarization modes in the magnetized plasma that
characterizes NS atmospheres are the
extraordinary mode (X-mode), which is mostly polarized perpendicular
to the $\veck-\vecB$ plane, and the ordinary mode (O-mode), which is
mostly polarized parallel to the $\veck-\vecB$ plane,
where $\veck$ specifies the direction of photon propagation
and $\vecB$ is the direction of the external magnetic field.
In our models, the temperature corrections
$\Delta T(\tau)$ at each Thomson depth $\tau$ are applied iteratively
until $\Delta T(\tau)/T(\tau)\la 1\%$, deviations from radiative
equilibrium are $\la 1\%$, and deviations from constant flux are
$\la 2\%$ (see \holaione\ for details of our numerical method).

For the high magnetic field models (Section~\ref{sec:himod}), we
include the effect of vacuum polarization.  As shown previously
(Lai \& Ho~2002, hereafter \laiho), vacuum polarization induces
conversion between
the low-opacity mode and the high-opacity mode, and the efficiency
of mode conversion depends on the photon energy.  In the absence of
a rigorous treatment of this effect, we follow \holaitwo\ by
considering the two limiting cases: complete mode conversion
(Section~\ref{sec:resultsvp}) and no mode conversion
(Section~\ref{sec:vpnc}).  Our treatment of the no mode conversion
case improves upon that given in \holaitwo.

A major complication when incorporating bound species in the atmosphere
models arises from the strong coupling between the center-of-mass motion
of the atom and the internal atomic structure.
\potcha\ have constructed thermodynamically
consistent EOS and opacity models of partially ionized
hydrogen plasmas in strong magnetic fields.
From the work of \potcha, we obtain tables for the
EOS and absorption opacities of hydrogen.  The EOS tables for
a given magnetic field are arranged by temperature and density.
The pressure at Thomson
depth $\tau$ ($=-\int\rho\keso\,dz$, where $\keso=\sigth/\mpr$)
is calculated from hydrostatic equilibrium.  With a given temperature
profile, we search the EOS tables for the nearest temperatures
and pressures and do a weighted average to obtain the density
profile.  We also extract the atomic fraction from the EOS tables.

The opacity tables contain the absorption opacity $\kabsa$
(where $\alpha=0,\pm$) as a function of temperature, density, and
photon energy.  
The absorption opacity for the modes ($j=1$ for X-mode, $j=2$ for
O-mode) is obtained from
\be
\kabsj=\sum_\alpha\polarevectwo\kabsa,
\ee
where $e_0^j=e_z^j$ is the z-component (along $\vecB$) of the mode
eigenvector and $e_\pm^j=(e_x^j\pm ie_y^j)/\sqrt{2}$ are the
circular components.
We calculate the polarization eigenvectors $\polarevecthree$
assuming that the medium is fully ionized since the
abundance of bound species is very low for the cases considered
here (see Section~\ref{sec:results});
this is not strictly correct, but the effect of a small amount
of bound species on the polarization of the medium is not known
at present (see
Bulik \& Pavlov~1996 for the case where the atmosphere is
completely neutral).\footnote{For all the models considered in
this paper, the neutral fraction is less than a few percent.}
For a given layer in the atmosphere model, which specifies a
particular temperature and density, the opacity $\kabsa(E)$ at a
given energy $E$ is obtained from the geometric average of the
nearest two or three energies of the table and used to compute
$\kabsj(E)$.  The scattering opacities are computed by the same
method as in \holaione\ and \holaitwo.

%%%%%%%%%%%%%%%%%%%%%%%%%%%%%%%%%%%%%%%%%%%%%%%%%%%%%%%%%%%%%%%%%%%
\section{Results} \label{sec:results}

%%%%%%%%%%%%%%%%%%%%%%%%%%%%%%%%%%%%%%%%%%%%%%%%%%%%%%%%%%%%%%%%%%%
\subsection{Low-Field ($B<\Bq$) Models} \label{sec:lowmod}

%-------------------------------------------
\begin{figure*}
\plotone{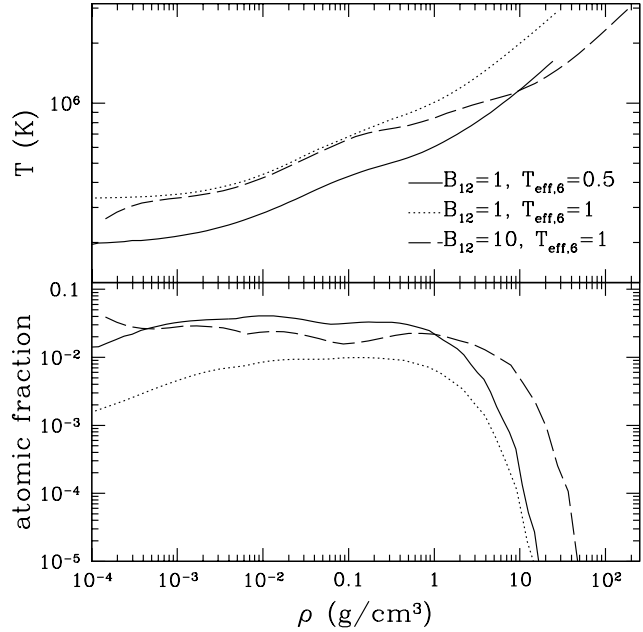}
\caption{
Temperature (upper panel) and atomic fraction (lower panel) as
functions of density for various partially ionized H
atmosphere models discussed in Section~\ref{sec:lowmod}:
magnetic field $B=10^{12}$~G and effective
temperature $\Teff=5\times 10^5$~K (solid lines),
$B=10^{12}$~G and $\Teff=10^6$~K (dotted lines), and
$B=10^{13}$~G and $\Teff=10^6$~K (dashed lines).
Atomic fraction is the number of H atoms with non-destroyed
energy levels divided by the total number of protons.
\label{fig:atmpro1213}
}
\end{figure*}
%-------------------------------------------

We consider three H atmosphere models in this subsection:
$B_{12}=1, \Teffsix=\Teff/(10^6\mbox{ K})=0.5$; $B_{12}=1, \Teffsix=1$;
$B_{12}=10, \Teffsix=1$.
Figure~\ref{fig:atmpro1213} shows the temperature and atomic fraction
profiles as a function of density.
The atomic fraction is the number of H atoms with non-destroyed
energy levels divided by the total number of protons (see \potcha).
The dependence of the atomic fraction on temperature and magnetic
field is evident; in particular, lower temperatures or higher
magnetic fields increase the abundance of bound species, with a
maximum of a few percent for the models shown here.

%-------------------------------------------
\begin{figure*}
\plotone{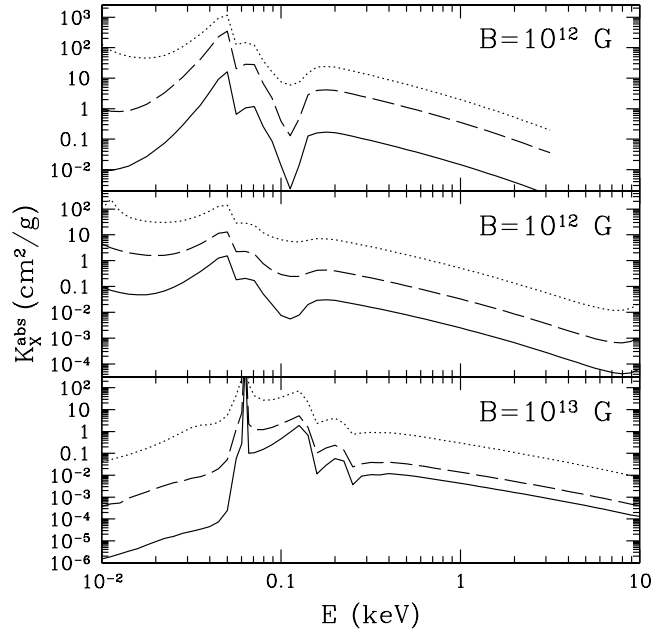}
\caption{
Angle-averaged X-mode absorption opacity $\Kabsx$ as a function
of energy for temperatures and densities representative of the
atmosphere models discussed in Section~\ref{sec:lowmod}.
The upper panel shows $\Kabsx$ appropriate for the model with
magnetic field $B=10^{12}$~G and effective
temperature $\Teff=5\times 10^5$~K ---
$T=4.2\times 10^5$~K and $\rho=0.093\mbox{ g cm$^{-3}$}$ (dotted line),
$T=2.4\times 10^5$~K and $\rho=4.0\times 10^{-3}\mbox{ g cm$^{-3}$}$
(dashed line),
$T=2.0\times 10^5$~K and $\rho=1.9\times 10^{-4}\mbox{ g cm$^{-3}$}$
(solid line).
The middle panel shows $\Kabsx$ appropriate for the model with
$B=10^{12}$~G and $\Teff=10^6$~K ---
$T=7.4\times 10^5$~K and $\rho=0.16\mbox{ g cm$^{-3}$}$ (dotted line),
$T=6.3\times 10^5$~K and $\rho=0.061\mbox{ g cm$^{-3}$}$ (dashed line),
$T=4.3\times 10^5$~K and $\rho=8.7\times 10^{-3}\mbox{ g cm$^{-3}$}$
(solid line).
The lower panel shows $\Kabsx$ appropriate for the model with
$B=10^{13}$~G and $\Teff=10^6$~K ---
$T=8.4\times 10^5$~K and $\rho=1.0\mbox{ g cm$^{-3}$}$ (dotted line),
$T=6.2\times 10^5$~K and $\rho=0.061\mbox{ g cm$^{-3}$}$ (dashed line),
$T=3.5\times 10^5$~K and $\rho=1.1\times 10^{-3}\mbox{ g cm$^{-3}$}$
(solid line).
The opacity curves in each panel are in descending order from hotter
and higher density to cooler and lower density, and the upper $\Kabsx$
(dotted) curves have been multiplied by 10 while the lower $\Kabsx$
(solid) curves have been divided by 10 for clarity.
See text for description of the opacity features.
\label{fig:opacity1213}
}
\end{figure*}
%-------------------------------------------

Figure~\ref{fig:opacity1213} shows the angle-averaged X-mode
absorption opacity $\Kabsx$, defined by
\be
\Kabsx = \frac{1}{4\pi}\int d\veck\,\kabsx(\veck).
\ee
as a function of energy for various temperatures and densities 
typical of the atmosphere models considered here.
Even though the bound species abundance is very low,
Fig.~\ref{fig:opacity1213} shows that they can still have a very
strong influence on the opacities.  Besides the proton cyclotron
line at $\Ebp=6.3B_{12}$~eV, there
are several strong features due to bound-bound and bound-free
transitions.  These are the $\bqnperp=0$ to $\bqnperp=1$ transition
at $E=51$~eV for $B_{12}=1$ and at $0.14$~keV for $B_{12}=10$,
the $\bqnperp=0$ to $\bqnperp=2$ transition at $E=75$~eV for
$B_{12}=1$ and at $0.23$~keV for $B_{12}=10$, and the bound-free
transition at $E=161$~eV for $B_{12}=1$ and at $0.31$~keV for $B_{12}=10$.
Because of magnetic broadening (see Section~\ref{sec:magatom}), these
features resemble bumps rather than ordinary spectral lines.

%-------------------------------------------
\begin{figure*}
\plotone{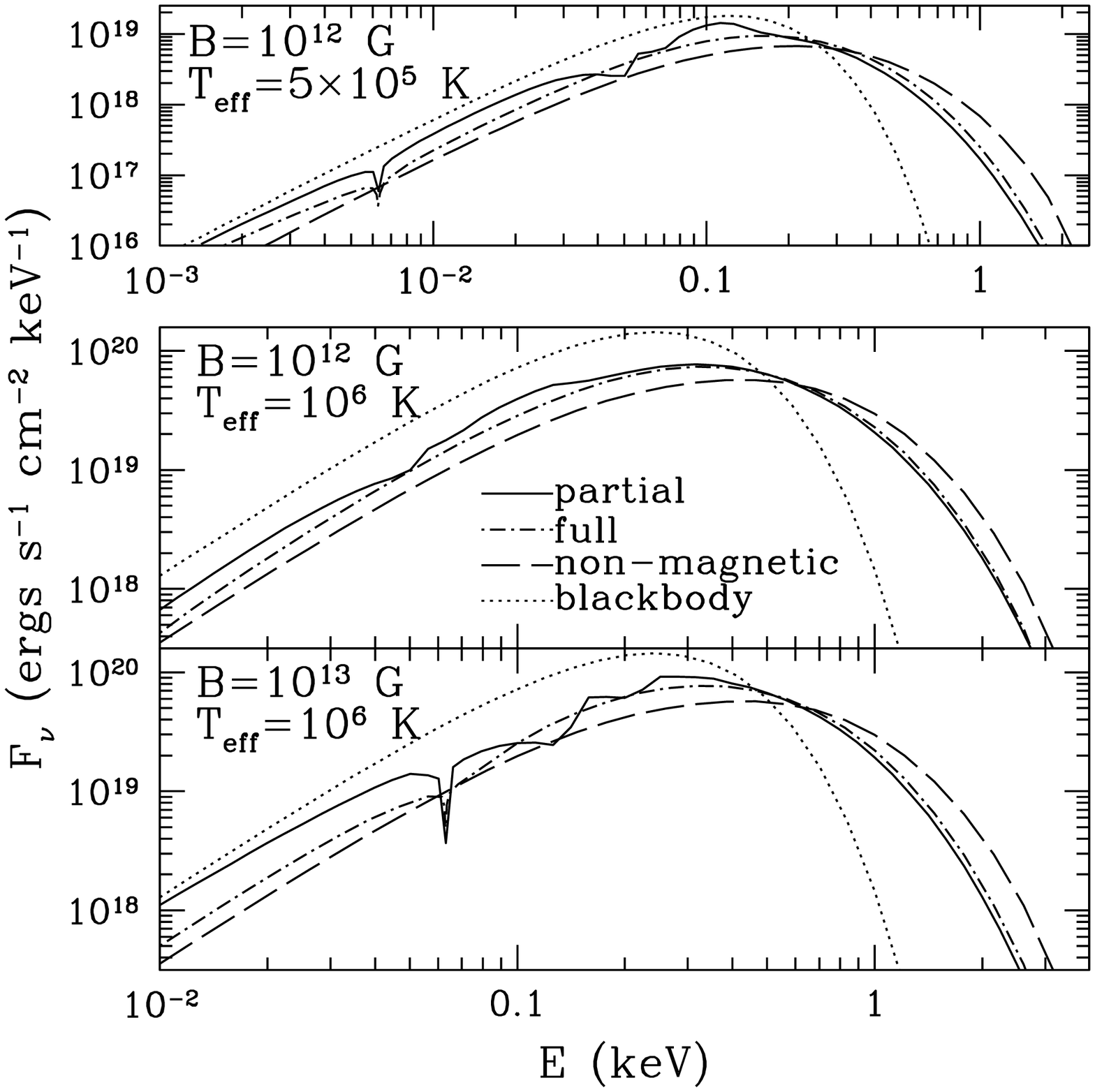}
\caption{
Spectra of hydrogen atmospheres with $B=10^{12}$~G and
$\Teff=5\times 10^5$~K (upper panel),
$B=10^{12}$~G and $\Teff=10^6$~K (middle panel),
and $B=10^{13}$~G and $\Teff=10^6$~K (lower panel).
The solid line is for a partially ionized atmosphere,
the dot-dashed line is for a fully ionized atmosphere,
the dashed line is for a fully ionized non-magnetic atmosphere,
and the dotted line is for a blackbody with $T=\Teff$.
The proton cyclotron line in the partially ionized spectra is
not resolved (see footnote~\ref{foot:ew}).
\label{fig:spectrum1213}
}
\end{figure*}
%-------------------------------------------

Figure~\ref{fig:spectrum1213} shows the spectra of the three atmosphere
models.  For comparison, we also show the spectra of the fully ionized
models at the same $B$ and $\Teff$, along with the non-magnetic fully
ionized models at the same $\Teff$ and the blackbody spectrum with $T=\Teff$.
The proton cyclotron line at $\Ebp=6.3\,B_{12}$~eV is clear in both the
fully ionized and partially ionized spectra.\footnote{
The width of the proton cyclotron line in all the partially ionized
models is due to the finite energy grid of the models and not an
indication of the true line width; the true width is narrower.
\label{foot:ew}}
At $B\la\Bq$, the effect of vacuum polarization on the atmosphere
structure and spectra is weak; therefore there is no suppression
of the proton cyclotron line (see Section~\ref{sec:himod} and \holaitwo),
and the proton cyclotron line can be very broad, especially when
$\Ebp\ga 3k\Teff$ (\holaione).
A comparison of the $B=10^{12}$~G models shows that spectral
features due to neutral species in the atmosphere are weaker
in the higher temperature model.  This is because the higher
temperatures decreases the abundance of neutral species
(see Fig.~\ref{fig:atmpro1213}) and thus reduces the effect of
the neutral species on the continuum opacity.

It is evident from Fig.~\ref{fig:spectrum1213}
that spectral features due to bound species at these field strengths
lie within the extreme UV
to very soft X-ray ($50\mbox{ eV}\la E\la 200$~eV) regime and
therefore are difficult to observe due to interstellar absorption.
However, the effect of bound species on the temperature profile
of the atmosphere and the continuum flux is significant.  In
particular, the optical flux is higher for the partially ionized
atmospheres compared to the fully ionized atmospheres.
On the other hand, the
partially and fully ionized models both yield very similar (neglecting
the spectral features) hard X-ray flux for a given effective temperature,
with the partially ionized model fluxes slightly lower than those
from the fully ionized models; thus fitting the observed hard X-ray flux
with fully ionized models would yield fairly accurate NS temperatures.

%%%%%%%%%%%%%%%%%%%%%%%%%%%%%%%%%%%%%%%%%%%%%%%%%%%%%%%%%%%%%%%%%%%
\subsection{Magnetar ($B>\Bq$) Models} \label{sec:himod}

When studying magnetar atmosphere models, it is important to include
the effect of vacuum polarization.  However, to clearly see the role
of bound species, we first consider models in which vacuum polarization
is neglected.

%%%%%%%%%%%%%%%%%%%%%%%%%%%%%%%%%%%%%%%%%%%%%%%%%%%%%%%%%%%%%%%%%%%
\subsubsection{No Vacuum Polarization} \label{sec:resultsnovp}

%-------------------------------------------
\begin{figure*}
\plotone{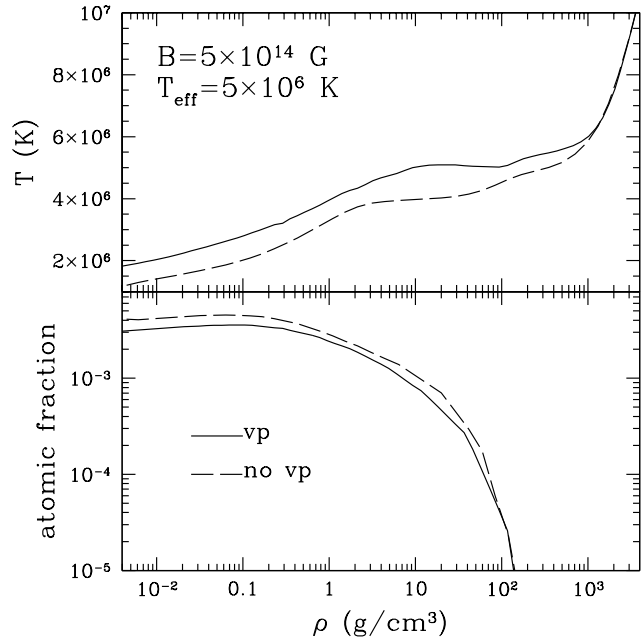}
\caption{
Temperature (upper panel) and atomic fraction (lower panel) as
functions of density for the partially ionized H
atmosphere models discussed in Section~\ref{sec:himod}:
$B=5\times 10^{14}$~G and $\Teff=5\times 10^6$~K with vacuum
polarization (solid lines) and without vacuum polarization (dashed lines).
Atomic fraction is the number of H atoms with non-destroyed
energy levels divided by the total number of protons.
\label{fig:atmprof14}
}
\end{figure*}
%-------------------------------------------

Figure~\ref{fig:atmprof14} shows the temperature and atomic fraction
profiles for the H atmosphere model with $B_{14}=B/(\mbox{10$^{14}$ G})=5$
and $\Teffsix=5$.
Here we use an extension of the \potcha\ EOS and opacity model
to $B > 10^{13.5}$~G. In this case, instead of the fits of
Potekhin~(1998) employed in \potcha\ (some of which would fail at
$B \ga 10^{13.5}$~G), we perform numerical calculations of the atomic
energies, sizes, and oscillator strengths following Potekhin~(1994).
(Some results of this extension of the \potcha\ model have been reported
in Chabrier, Douchin, \& Potekhin~2002.)
Though the magnetic field increases the binding energy of
the hydrogen atom to $> 0.5$~keV, the atomic abundance remains
$< 1\%$ in the atmospheric layers.

%-------------------------------------------
\begin{figure*}
\plotone{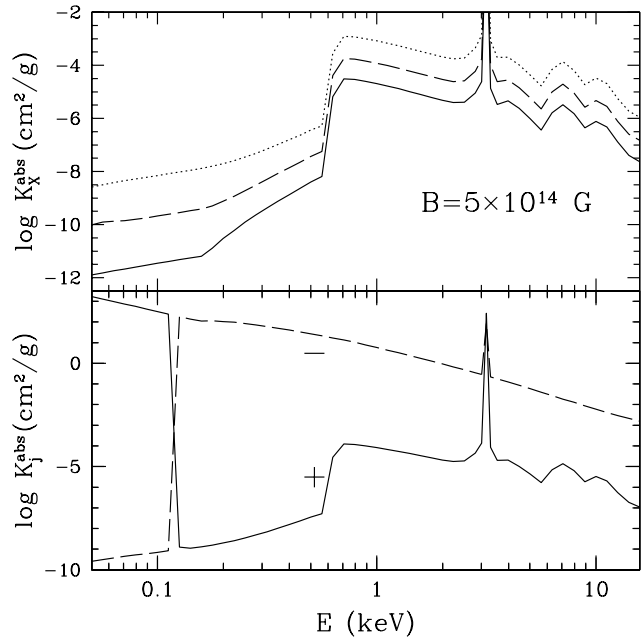}
\caption{
Angle-averaged absorption opacities as functions
of energy for temperatures and densities representative of the
$B=5\times 10^{14}$~G and $\Teff=5\times 10^6$~K
atmosphere model including vacuum polarization
discussed in Section~\ref{sec:himod}.
The upper panel shows the X-mode opacity $\Kabsx$ at
$T=2.9\times 10^6$~K and $\rho=0.13\mbox{ g cm$^{-3}$}$ (dotted line),
$T=2.4\times 10^6$~K and $\rho=0.031\mbox{ g cm$^{-3}$}$ (dashed line),
$T=1.8\times 10^6$~K and $\rho=4.0\times 10^{-3}\mbox{ g cm$^{-3}$}$
(solid line).
The opacity curves in the upper panel are in descending order from hotter
and more dense to cooler and lower density, and the upper $\Kabsx$
(dotted) curves have been multiplied by 10 while the lower $\Kabsx$
(solid) curves have been divided by 10 for clarity.
The lower panel shows the plus-mode and minus-mode opacities $\Kabsj$
($j=\pm$) and $T$ and $\rho$ corresponding to the dotted
curve in the upper panel;
using $\Kabsp$ and $\Kabsm$ is appropriate for models in which complete
mode conversion is assumed (see Section~\ref{sec:resultsvp}).
See text for description of the opacity features.
\label{fig:opacity14}
}
\end{figure*}
%-------------------------------------------

The upper panel of Figure~\ref{fig:opacity14} shows the
angle-averaged X-mode opacity $\Kabsx$ as a function of energy
for various temperatures and densities typical of the high magnetic
field model considered here.
Besides the proton cyclotron line at $\Ebp=0.63\,B_{14}$~keV,
the sharp change in opacity at
$0.76$~keV for $B_{14}=5$ is due to photoionization.
The features at $E>\Ebp$ are the partial photoionization
edges for transitions to the continuum states with $\bqnperp > 0$,
which are merged with autoionization resonances just below the corresponding
partial thresholds and extended redwards due to magnetic broadening
(see Potekhin \& Pavlov~1997).  The bound-bound transitions are not
seen in Fig.~\ref{fig:opacity14} because the bound states with
$\bqnperp > 0$ merge with the continuum at such field strengths
because of the $\bqnperp\hbar\omegabp$-term in $E_\bqnperp$.

%-------------------------------------------
\begin{figure*}
\plotone{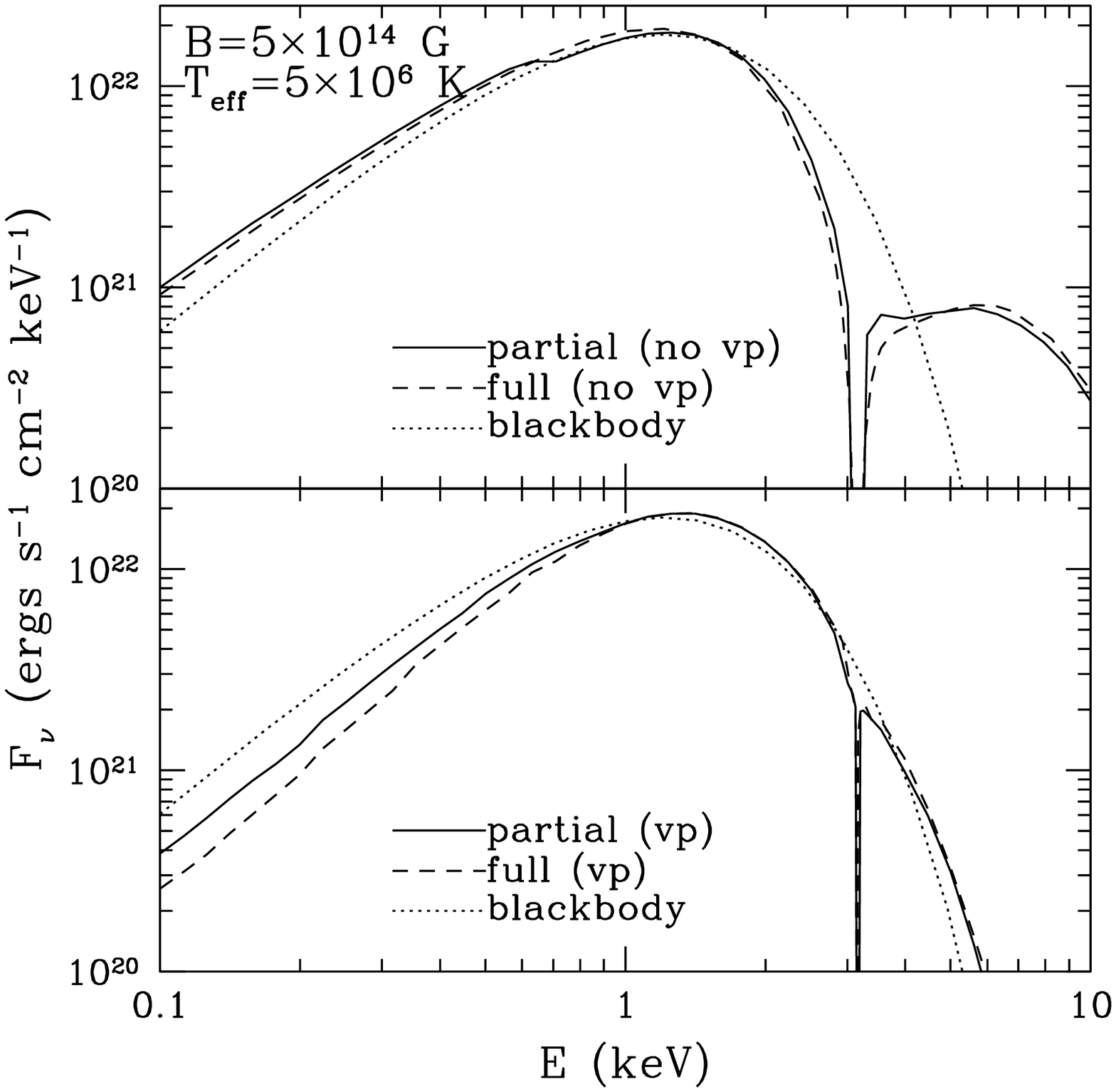}
\caption{
Spectra of hydrogen atmospheres with $B=5\times 10^{14}$~G and
$\Teff=5\times 10^6$~K.
The upper panel shows the partially ionized (solid line) and
fully ionized (dashed line) atmospheres with no vacuum polarization.
The lower panel shows the partially ionized (solid line) and
fully ionized (dashed line) atmospheres with vacuum polarization
and assuming complete mode conversion.
The dotted lines are for a blackbody with $T=5\times 10^6$~K.
The proton cyclotron line in the partially ionized spectra is
not resolved (see footnote~\ref{foot:ew}).
\label{fig:spectrum514}
}
\end{figure*}
%-------------------------------------------

Figure~\ref{fig:spectrum514} shows the spectra of the models
with $B=5\times 10^{14}$~G and $\Teff=5\times 10^6$~K.
The upper panel is a comparison of partially and fully ionized
models with no vacuum polarization.
Broad absorption features at $E\sim 0.76$~keV and 4~keV are due to
bound-free transitions to different continuum states (the latter
feature is from the blending of two magnetically broadened partial
photoionization edges at 3.4~keV and 6.5~keV;
see Fig.~\ref{fig:opacity14}),
while the continuum flux between the models is not significantly different.
Also, as noted in \holaione\ and \holaitwo\ (see also Zane \etal~2001),
the proton cyclotron line is very broad when vacuum polarization is
neglected.

%%%%%%%%%%%%%%%%%%%%%%%%%%%%%%%%%%%%%%%%%%%%%%%%%%%%%%%%%%%%%%%%%%%
\subsubsection{Including Vacuum Polarization: Complete Mode Conversion}
\label{sec:resultsvp}

The main effect of vacuum polarization on radiative transfer is
the ``vacuum resonance,'' which occurs when the effects of the
vacuum and plasma on the polarization modes cancel each other.
For a photon of energy $E(=\hbar\omega)$, the vacuum resonance occurs at
density
\be
\densvp\approx 0.96\,\Ye^{-1}E_1^2B_{14}^2f(B)^{-2}\mbox{ g cm$^{-3}$},
\label{eq:densvp}
\ee
where $E_1=E/(\mbox{1 keV})$ and $f(B)$ is a slowly-varying function
of $B$ of order unity.
The two photon modes are almost linearly polarized away from the
resonance, while they both become circularly polarized near $\densvp$.
\laiho\ showed that a photon propagating in the atmospheric
plasma can adiabatically convert from the low-opacity X-mode to the
high-opacity O-mode (and vice versa) as it crosses $\densvp$.
For this conversion to be effective, the adiabatic condition
\be
E_{\rm ad}\ga 1.49\,\lb f(B)\tan\theta_B|1-u_{\rm i}|\rb^{2/3}
\lp 5\mbox{ cm}/H_\rho\rp^{1/3} \mbox{keV},
\label{eq:enadiabat}
\ee
where $\theta_B$ is the angle between $\veck$ and $\vecB$,
$u_{\rm i}=\omegabp^2/\omega^2$, and $H_\rho=|dz/d\ln\rho|$ is the
density scale height along the ray, must be satisfied.
Clearly, resonant mode conversion can significantly affect
radiative transfer (\laiho; \holaitwo; Lai \& Ho~2003).

To rigorously treat the radiative transfer across the vacuum
resonance, one must go beyond the modal description of the
radiation field by solving the transfer equation in terms of
the photon intensity matrix (Lai \& Ho~2003).  In the present
paper, as in \holaitwo, we consider two limiting cases.  In
the ``complete mode conversion'' case, we assume photons of all
energies change their mode characteristics (from X to O-mode or
vice versa) when crossing the vacuum resonance.  This can be
achieved by using the plus-mode and minus-mode basis ($j=\pm$) in
the radiative transfer equation (see \holaitwo\ for details).
At a given density $\rho$, the characteristics of the $\pm$-modes
switch at the energy
\be
\Evp\approx 1.02\,\Ye^{1/2}\rho_1^{1/2}B_{14}^{-1}f(B)\mbox{ keV},
\label{eq:evp}
\ee
where $\rho_1=\rho/(\mbox{1 g cm$^{-3}$})$.

Figure~\ref{fig:atmprof14} shows the temperature and atomic fraction
profiles for the partially ionized model which includes vacuum
polarization in the complete mode conversion limit.
As noted in \holaitwo, vacuum polarization increases the temperature
at each depth relative to the case without vacuum polarization.
The higher temperatures result in lower atomic fractions for a given
magnetic field.

The lower panel of Fig.~\ref{fig:opacity14} shows the angle-averaged
opacity $\Kabsj$ for $j=\pm$ as a function of energy at
$\rho=0.13\mbox{ g cm$^{-3}$}$ and $T=2.9\times 10^6$~K, which are
characteristic of the $B=5\times 10^{14}$~G atmosphere.
The distinction between the plus and minus-modes ($j=\pm$) and the
X and O-modes is discussed in \holaitwo\ (in particular, see Fig.~4
of that work).
At this particular temperature and density, the plus-mode (minus-mode)
is in the high (low) opacity state for $E<\Evp$ but manifests as
the low (high) opacity state for $E>\Evp$.

Fig.~\ref{fig:spectrum514} compares the spectra of different models
with and without vacuum polarization.
First, we see that, with vacuum polarization, the continuum flux at
high energies for the partially ionized model is very similar to the fully
ionized model [for the parameters shown in Fig.~\ref{fig:spectrum514},
$F_\nu\mbox{(partial)}/F_\nu\mbox{(full)}<1.5$ at $E>0.1$~keV].
Second, vacuum polarization softens the high energy tail of the
spectrum in both the partially ionized and fully ionized cases,
although the emission is still harder than the blackbody spectrum
because of the non-grey opacities.
The softer spectrum is due to the density dependence of the vacuum resonance
energy $\Evp$ and the large density gradient present in the atmosphere;
the change in opacity at the vacuum resonance causes
the high energy photons emerging from the atmosphere to be
produced from a shallower, cooler layer in the atmosphere than
in the case when vacuum polarization is neglected
(see Fig.~2 of Ho \etal~2003).
Third, as discussed in \holaitwo, this vacuum-induced depression
of continuum flux strongly suppresses the proton cyclotron line
by making the photospheres inside and outside the line more
similar (see also footnote~\ref{foot:ew}).  We now see that
vacuum polarization also suppresses spectral features due
to the bound species of hydrogen.  The reduced width of the proton
cyclotron line and spectral features associated with bound transitions
makes these features difficult to observe with current X-ray
detectors (see Section~\ref{sec:discussion}).

%%%%%%%%%%%%%%%%%%%%%%%%%%%%%%%%%%%%%%%%%%%%%%%%%%%%%%%%%%%%%%%%%%%
\subsubsection{Including Vacuum Polarization: No Mode Conversion}
\label{sec:vpnc}

The opposite limit for treating the vacuum polarization effect
is to assume no mode conversion across $\densvp$.  This is achieved by
using the X-mode and O-mode as the mode basis in the transfer equation
(see \holaitwo).  For this case, the X-mode opacity has a much lower
opacity than the O-mode opacity
except at the resonance peak $E=\Evp$ (where $\kappa_X=\kappa_O$).
As discussed in \laiho, even without
mode conversion, the optical depth across this opacity feature can be
significant; therefore, the X-mode radiative transfer is affected.

%-------------------------------------------
\begin{figure*}
\plotone{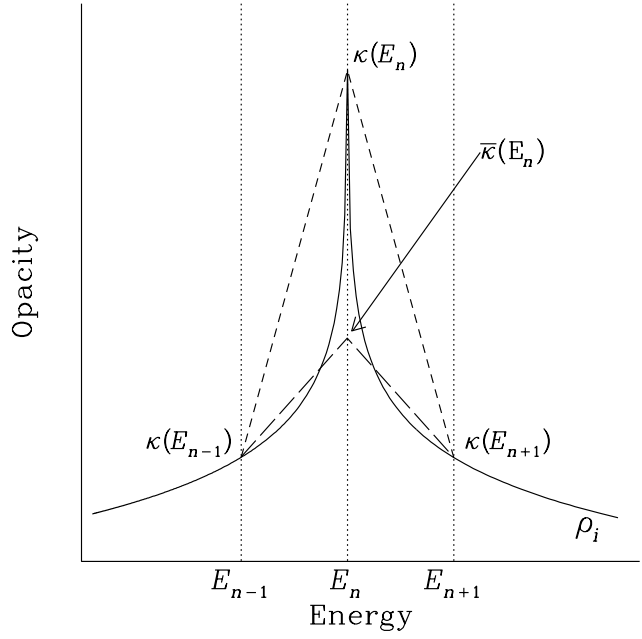}
\caption{
Schematic picture of the radiative opacity as a function of energy at
the density grid point $\rho_i$.  The vacuum resonance feature occurs
at higher energies at larger densities (depths)
[see eq.~(\ref{eq:evp})].
In the adaptive energy grid method of \holaitwo\ and
Section~\ref{sec:vpnc}, the energy grid points $\Ena, \Enc, \Enb$
are chosen to be equal to the vacuum resonance energy at one of
the density grid points, i.e., $\Enc=\Evp(\rho_i)$.
The short-dashed line indicates the apparent opacity obtained by
interpolating energy grid points $\Ena, \Enc, \Enb$,
while the long-dashed line indicates the apparent opacity obtained by
the method described in Section~\ref{sec:vpnc}.
\label{fig:vpnc}
}
\end{figure*}
%-------------------------------------------

We present here improvements made to our models with no mode conversion.
The atmosphere models contain energy grid points $\Enc$ and
density grid points $\rho_i$.
As described in \holaitwo, we use
an adaptive energy grid to account for the narrow, density-dependent
opacity feature at $\Evp$, i.e., an equal number of density and
energy grid points is used with
every energy grid point being placed at $\Enc=\Evp(\rho_i)$.
We noted in \holaitwo, however, that this implementation
overestimates the strength of the opacity feature on the
atmosphere structure and spectra.
This is illustrated in Figure~\ref{fig:vpnc}:
The area under the short-dashed line overestimates the area under
the solid line, which represents the true opacity.

To rectify this problem, we introduce
an effective opacity at the resonance, $\bar{\kappa}(\Evp)$,
given by
\begin{eqnarray}
\bar{\kappa}(\Enc=\Evp) & = & \frac{2}{\Enb-\Ena}
 \lb\int_{\Ena}^{\Enb}\kappa(E)\,dE \right. \nonumber \\
 && - \frac{1}{2}\kappa(\Ena)\lp\Enc-\Ena\rp \nonumber \\
 && \left. - \frac{1}{2}\kappa(\Enb)\lp\Enb-\Enc\rp\rb, \label{eq:kvpeff}
\end{eqnarray}
where the integral is calculated using an auxiliary energy grid that
has much finer resolution.  Using $\bar{\kappa}(\Evp)$ preserves
the equivalent width of the resonance feature: In Fig.~\ref{fig:vpnc},
the area under the long-dashed line is equal to the area under
the solid line.
As would be expected, atmosphere models based on this improved method
but with different numbers of energy grid points yield very similar
results.

%-------------------------------------------
\begin{figure*}
\plotone{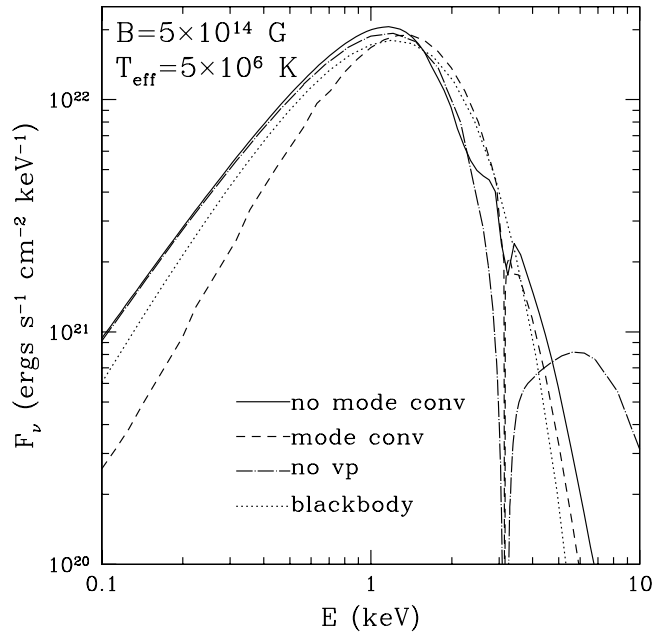}
\caption{
Spectra of fully ionized hydrogen atmospheres with
$B=5\times 10^{14}$~G and $\Teff=5\times 10^6$~K.
The solid line and dashed line are the atmospheres with vacuum
polarization but no mode conversion and complete mode conversion,
respectively, the dot-dashed line is the atmosphere without vacuum
polarization, and
the dotted line is for a blackbody with $T=5\times 10^6$~K.
\label{fig:spectrumnc}
}
\end{figure*}
%-------------------------------------------

The spectra for fully ionized hydrogen atmosphere models with
$B=5\times 10^{14}$~G and $\Teff=5\times 10^6$~K is shown in
Figure~\ref{fig:spectrumnc}.
Note that the width of the proton cyclotron line in the (no mode
conversion) adaptive energy grid models is not the true width of
the line.
The broad depression in the spectrum due to the density
dependence of the vacuum resonance feature is
shifted to higher energies since the old calculation (see \holaitwo)
overestimated the effect of the resonance feature.
The two limiting cases of complete mode conversion and no mode
conversion should approximately bracket the true spectrum.
We also note that the no mode conversion flux
spectrum should be suppressed at low energies due to the dense
plasma effect described in Section~\ref{sec:plasma}, and thus the
two limiting cases should have more similar fluxes at lower
energies than is shown in Figure~\ref{fig:spectrumnc}.

%%%%%%%%%%%%%%%%%%%%%%%%%%%%%%%%%%%%%%%%%%%%%%%%%%%%%%%%%%%%%%%%%%%
\subsubsection{Proper Treatment of Free-Free Opacity}
\label{sec:opff}

As discussed in \potcha, the traditional formula (see Pavlov \etal~1995)
for the opacity due to free-free absorption of both
electrons and ions given by
\be
\kappa_\alpha^{\rm ff} \propto \frac{1}{\lp\omega+\alpha\omegab\rp^2}
 + \lp\frac{\me}{\mpr}\rp^2 \frac{1}{\lp\omega-\alpha\omegabp\rp^2}
\qquad \mbox{(incorrect)},
\label{eq:opffwrong}
\ee
where $\alpha=0,\pm 1$ and $\hbar\omegab=11.6\,B_{12}$~keV
is the electron cyclotron energy,
misses an interference effect due to electron-ion collisions.
The correct free-free absorption opacity is given by
(see, e.g., PC for more discussion)
\be
\kappa_\alpha^{\rm ff} \propto \frac{\omega^2}{\lp\omega+\alpha\omegab\rp^2
\lp\omega-\alpha\omegabp\rp^2} \qquad \mbox{(correct)}.
\label{eq:opff}
\ee
When $\omega\gg\omegabp$, equations~(\ref{eq:opffwrong}) and (\ref{eq:opff})
are very nearly the same.
However, when $\omega<\omegabp$, the correct opacity is suppressed
by a factor $\sim \omega^2/\omegabp^2$ compared to the incorrect one.
Thus we expect that, in the atmosphere models using the correct
opacity given by equation~(\ref{eq:opff}), the photons with
$E<\Ebp$ will be generated from deeper, hotter layers as opposed
to models which use the incorrect opacity given by
equation~(\ref{eq:opffwrong}).
All previous atmosphere models
adopted equation~(\ref{eq:opffwrong}) and are therefore problematic if
photons with $E<\Ebp$ are important (see, however, Lloyd~2003).

%-------------------------------------------
\begin{figure*}
\plotone{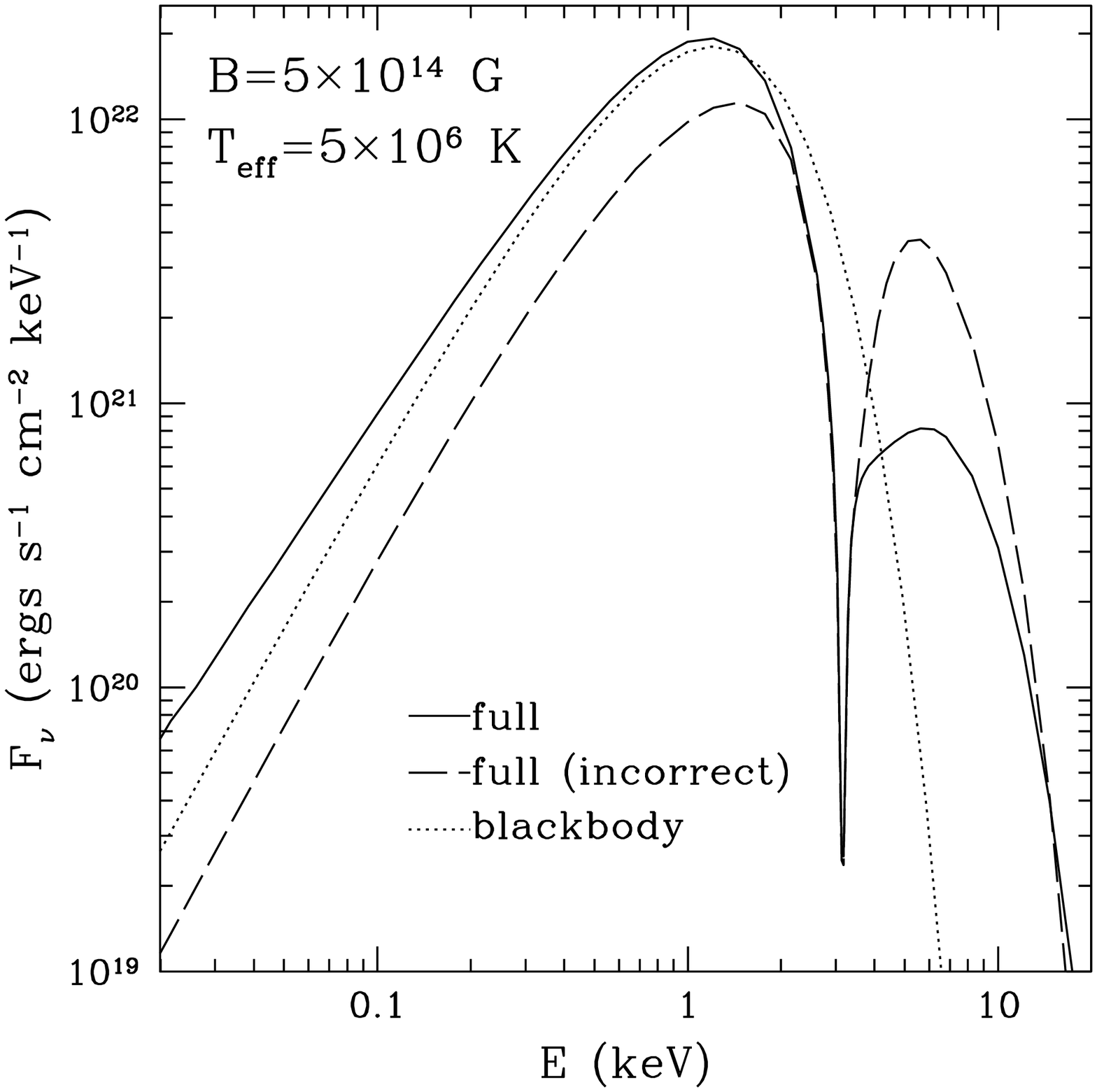}
\caption{
Spectra of fully ionized hydrogen atmospheres with $B=5\times 10^{14}$~G and
$\Teff=5\times 10^6$~K using the correct (solid) expression for the
free-free absorption opacity [eq.~(\ref{eq:opff})] versus the
incorrect (dashed) expression [eq~(\ref{eq:opffwrong})].
The dotted line is for a blackbody with $T=5\times 10^6$~K.
\label{fig:spectrumwrong}
}
\end{figure*}
%-------------------------------------------

Figure~\ref{fig:spectrumwrong} compares atmosphere spectra generated
using the correct expression for the free-free absorption opacity
[eq.~(\ref{eq:opff})] and using the incorrect expression
[eq.~(\ref{eq:opffwrong})].
The model shown here is a more extreme
example since $\Ebp\ga k\Teff$ (in order to maintain constant total
flux, a large redistribution of the specific energy flux occurs).
When $\Ebp\ll k\Teff$, such as for the low magnetic field cases
considered in Section~\ref{sec:lowmod}, the difference between
using the correct expression versus the incorrect one is small
(except possibly in the region $\omega<\omegabp$; but this region
of the spectrum carries negligible flux).
All models considered in this paper use the correct expression.

%%%%%%%%%%%%%%%%%%%%%%%%%%%%%%%%%%%%%%%%%%%%%%%%%%%%%%%%%%%%%%%%%%%
\section{Dense Plasma Effect} \label{sec:plasma}

\begin{figure*}
\plotone{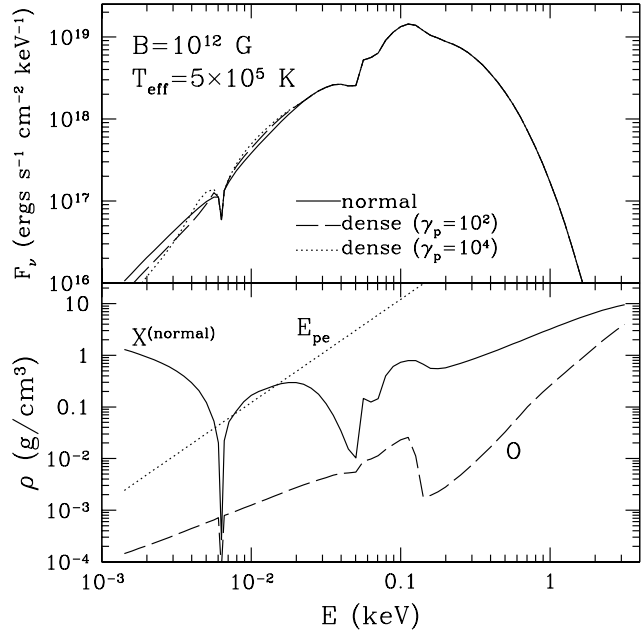}
\caption{
Spectra (upper panel) and density at the photosphere (lower panel)
for partially ionized hydrogen atmospheres with $B=10^{12}$~G
and $\Teff=5\times 10^5$~K.
In the upper panel, the solid line shows the spectrum (``normal'')
when the dense plasma effect discussed in Section~\ref{sec:plasma}
is neglected, while the dashed line and dotted line show the spectra
(``dense'') when the dense plasma effect is approximately accounted
for using $\gamma_{\rm p}=10^2$ and $10^4$, respectively.
The lower panel shows the density at the X-mode photon decoupling
layer (at which $\tau_\nu\approx 1$; solid line)
and the O-mode photon decoupling layer (dashed line).
The dotted line denotes the density at which the photon frequency
equals the plasma frequency, i.e.,
$E=\hbar\omegap=28.71\,\rho_1^{1/2}$~eV.
\label{fig:spectrumplasma}
}
\end{figure*}

Here we comment on a main uncertainty associated with atmosphere
models of strongly magnetized NSs.
Although we have included the medium effect on the polarization modes,
at sufficiently high densities or low photon energies, the refractive
index of the medium deviates appreciably from unity.
In particular, at the photon decoupling layer in the atmosphere
(where the optical depth $\tau_\nu\approx 1$), the plasma frequency
$\omegap$ may exceed the photon frequency $\omega$.
For example, in the $B=10^{12}$~G and $\Teff=5\times 10^5$~K model
shown in Section~\ref{sec:lowmod}, the X-mode photons (which are
the main carriers of the energy flux) with $E\la 10$~eV decouple
from the atmosphere layer where
$E<\hbar\omegap=28.71\,(\rho/\mbox{1 g cm$^{-3}$})^{1/2}$~eV
(see lower panel of Fig.~\ref{fig:spectrumplasma}).
Clearly we expect that the actual emission at these energies is
lower than what is shown in our model spectrum
(see upper panel of Fig.~\ref{fig:spectrum1213}).

%-------------------------------------------
%-------------------------------------------

A quantitative study of this dense plasma effect is beyond the scope of
this paper.  However, we can obtain an estimate of its effect by a
simple procedure.  In the atmosphere model, we evaluate the parameter
$\vel(=\omegap^2/\omega^2)$; whenever $\vel>1$, we artificially
increase all the opacities at that energy and depth by a factor
$\gamma_{\rm p}(\gg 1)$.  This simulates the suppression of radiation
below the plasma frequency cutoff.  The resulting spectrum is shown in
Fig.~\ref{fig:spectrumplasma}, where we have chosen $\gamma_{\rm p}=10^2$
and $10^4$ and labeled the results as ``dense.''
The result of Section~\ref{sec:lowmod}, where we ``blindly'' carry out
the calculations even when $\vel>1$, is labeled as ``normal'' in
Fig.~\ref{fig:spectrumplasma}.
We see that the spectrum is largely unaffected since $\vel<1$ at
most energies.
However, we see the flux at low energies ($E\la 5$~eV)
is suppressed [with $F_\nu^{\rm (norm)}/F_\nu^{\rm (dense)}\approx 1.4$
at 2~eV for $\gamma_{\rm p}=10^2$ and
$F_\nu^{\rm (norm)}/F_\nu^{\rm (dense)}\approx 1.7$ at 2~eV for
$\gamma_{\rm p}=10^4$].
The slight increase in flux at $\sim$7-20~eV is due to flux
redistribution to maintain constant total flux or radiative equilibrium.

For the magnetar models, the strong suppression of the X-mode opacities
by the magnetic field causes the photons to be generated from deeper,
denser atmosphere layers.
Therefore, the dense plasma effect may be even more important.
(However, this can be offset by vacuum polarization which makes the
photosphere shallower; see Section~\ref{sec:himod}.)
Clearly this dense plasma effect needs to be understood better before
accurate comparisons can be made with, for example, optical and infrared
observations.

%%%%%%%%%%%%%%%%%%%%%%%%%%%%%%%%%%%%%%%%%%%%%%%%%%%%%%%%%%%%%%%%%%%
\section{Discussion} \label{sec:discussion}

We have presented results of the first magnetic hydrogen atmosphere
models of isolated neutron stars based on the best current equation
of state and opacities of magnetized, partially ionized H plasmas
(\potcha).  For models with ``ordinary'' magnetic field strengths
($10^{12}$~G$\la B\la\Bq=4.4\times 10^{13}$~G),
the spectral lines associated with bound species lie in the extreme
UV to very soft X-ray energy bands and are difficult to observe
due to interstellar absorption at these energies
(though a broad absorption feature at $\sim 0.2$~keV was seen
in RBS~1223; Haberl \etal~2003; see also Ho \& Lai~2003b).  However, the
opacities are sufficiently different from the fully ionized
opacities that they can change the atmosphere structure and
continuum flux.

We find that, for the higher magnetic field models ($B>\Bq$),
vacuum polarization suppresses both the proton cyclotron line
and spectral lines due to the bound species.  As a result, the
thermal spectra are almost featureless and blackbody-like.
Recent observations of several AXPs with {\it Chandra} and
{\it XMM-Newton} X-ray telescopes failed to resolve any
significant line features in the spectra
(e.g., Patel \etal~2001, 2003; Juett \etal~2002; Tiengo \etal~2002;
Morii \etal~2003);
thus this could possibly be indicating the presence of vacuum
polarization effects in these magnetar candidates.

Our results may also be useful in understanding spectral observations 
of dim isolated neutron stars, such as RX~J$1856.5-3754$ and
RX~J$0720.4-3125$ (see, e.g., Pavlov \etal~2002; Kaplan \etal~2003).
Since $k\Teff\sim\mbox{tens of}$~eV for these neutron stars, the
presence of neutral atoms must be accounted for in the atmosphere
models if $B\ga 10^{12}$~G.
We find that one-temperature partially ionized hydrogen atmosphere
models improve the fitting of the X-ray data (compared to fully
ionized models), although they still do not provide a satisfactory
fit to the combined X-ray and optical spectra.  One possible
solution is ``thin'' atmosphere models in which a gaseous hydrogen
atmosphere lies on top of a condensed matter (see Motch \etal~2003).
We find that the spectra of the neutron stars RX~J$1856.5-3754$
and RX~J$0720.4-3125$ can be crudely fit by our $B=10^{12}$~G
and $\Teff\sim 5\times 10^5$~K partially ionized hydrogen ``thin''
atmosphere model (Ho \etal, in preparation).
The resulting optical spectrum of the model only overpredicts the
optical data by $\la 1.5$.
Furthermore, when the dense plasma effect discussed in
Section~\ref{sec:plasma} is taken into account, the suppression
of low energy flux may lead to an even better fit of the optical spectrum.

{\it Chandra} and {\it XMM-Newton}
observations revealed absorption features at 0.7 and 1.4~keV
in the spectra of the neutron star 1E~$1207.4-5209$
(Mereghetti \etal~2002b; Sanwal \etal~2002).  Sanwal \etal~(2002)
contend the features are due to atomic transitions of
singly-ionized helium in the neutron star atmosphere with a
superstrong magnetic field, $B\approx 1.5\times 10^{14}$~G.
Hailey \& Mori~(2002) and Mori \& Hailey~(2003) attribute the features
to transitions of mid-$Z$ oxygen or neon in a $\sim 10^{12}$~G
magnetic field, while Mereghetti \etal~(2002b) interpret the
features to be the result of high-$Z$ metals at $\sim 10^{12}$~G.
If 1E~$1207.4-5209$ is indeed a neutron star with superstrong
magnetic field ($B\ga\Bq$) and the properties of the helium atom
are similar to those of the hydrogen atom at these field strengths,
then our results show that any spectral features would not be resolved
due to their suppression by vacuum polarization.
However, we cannot be more definitive because
of the unknown properties of helium at these field strengths and,
as discussed in Section~\ref{sec:nummeth}, we assumed that the presence
of a very small amount of neutral atoms has no effect on the
polarization properties of the medium.
Furthermore, recent observations have revealed an additional
feature at 2.1~keV and possibly another at 2.8~keV thus hinting
at an electron cyclotron origin (with $B=8\times 10^{10}$~G)
for these features (Bignami \etal~2003).

Several caveats/uncertainties are worth mentioning.
(1) The radiative transfer formalism adopted in
our work relies on the transfer of two photon polarization modes.
This is inadequate for treating the vacuum polarization-induced
resonant mode conversion effect, especially because the
effectiveness of mode conversion depends on photon energy
(see \laiho; \holaitwo).
To properly account for the mode conversion effect, as well as mode
collapse and the breakdown of Faraday depolarization, one must go
beyond the modal description of the radiation field by formulating
and solving the transfer equation in terms of the photon intensity matrix
(or Stokes parameters) and including the effect of a nontrivial
refractive index (see Lai \& Ho~2003).
(2) As discussed in Section~\ref{sec:plasma}, the dense plasma effect
on radiative transfer can be important, especially for low energy photons
and high magnetic fields.  These issues require further study.

\acknowledgments

We thank Marten van Kerkwijk for useful discussions and for also
obtaining rough fits of our
models to the spectra of RX~J$1856.5-3754$ and RX~J$0720.4-3125$.
We thank the anonymous referee for useful comments.
We are grateful to the Cornell Hewitt Computer Laboratory for the
use of its facilities.
This work is supported in part by NASA grant
NAG 5-12034 and NSF grant AST 0307252.
A.Y.P. acknowledges the hospitality of the Cornell Astronomy
Department, where part of this work was performed.
The work of A.Y.P. is supported by RFBR grants 02-02-17668
and 03-07-90200.

%%%%%%%%%%%%%%%%%%%%%%%%%%%%%%%%%%%%%%%%%%%%%%%%%%%


\begin{thebibliography}{longestkeymustbeshorterthanthis99}

\bibitem[Adler(1971)]{adler71} Adler, S.L. 1971, Ann. Phys., 67, 599
\bibitem[Becker(2000)]{becker00} Becker, W. 2000, in Proc. IAU Symp. 195,
 Highly Energetic Physical Processes and Mechanisms for Emission
 from Astrophysical Plasmas, eds. Martens, P.C.H., Tsuruta, S., \&
 Weber, M.A. (San Francisco: ASP), p.49
\bibitem[Bignami\etal(2003)]{bignamietal03} Bignami, G.F., Caraveo, P.A.,
 De Luca, A., \& Mereghetti, S. 2003, Nature, 423, 725
\bibitem[Bulik \& Pavlov(1996)]{bulikpavlov96} Bulik, T. \& Pavlov, G.G. 1996,
 ApJ, 469, 373
\bibitem[Chabrier \etal(2002)]{chabrieretal02} Chabrier, G., Douchin, F., \&
 Potekhin, A.Y. 2002, J. Phys.: Condens. Matter, 14, 9133
\bibitem[Gnedin \etal(1978)]{gnedinetal78} Gnedin, Yu.N., Pavlov, G.G.,
 \& Shibanov, Yu.A. 1978, Sov. Astron. Lett., 4, 117
\bibitem[Haberl \etal(2003)]{haberletal03} Haberl, F., Schwope, A.D.,
 Hambaryan, V., Hasinger, G., \& Motch, C. 2003, A\&A, 403, L19
\bibitem[Hailey \& Mori(2002)]{haileymori02} Hailey, C.J. \& Mori, K. 2002,
 ApJL, 578, L133
\bibitem[Heyl \& Hernquist(1997)]{heylhernquist97} Heyl, J.S. \&
 Hernquist, L. 1997, Phys. Rev. D, 55, 2449
\bibitem[Ho \& Lai(2001)]{holai01} Ho, W.C.G. \& Lai, D. 2001,
 MNRAS, 327, 1081 (\holaione)
\bibitem[Ho \& Lai(2003a)]{holai03a} Ho, W.C.G. \& Lai, D. 2003a,
 MNRAS, 338, 233 (\holaitwo)
\bibitem[Ho \& Lai(2003b)]{holai03b} Ho, W.C.G. \& Lai, D. 2003b,
 ApJ, to be submitted
\bibitem[Ho \etal(2003)]{hoetal03} Ho, W.C.G., Lai, D., Potekhin, A.Y.,
 \& Chabrier, G. 2003, Adv. Sp. Res., in press (astro-ph/0212077)
\bibitem[Hurley(1999)]{hurley99} Hurley, K. 2000, in Gamma-Ray
 Bursts: Fifth Huntsville Symposium, AIP Conf. Proc. 526, eds.
 Kippen, R.M., Mallozzi, R.S., \& Fishman, G.J. (New York: AIP), p.515
\bibitem[Juett \etal(2002)]{juettetal02} Juett, A.M., Marshall, H.L.,
 Chakrabarty, D., \& Schulz, N.S. 2002, ApJL, 568, L31
\bibitem[Kaplan \etal(2003)]{kaplanetal03} Kaplan, D.L., Kulkarni, S.R.,
 \& van Kerkwijk, M.H. 2003, ApJL, 588, L33
\bibitem[Lai(2001)]{lai01} Lai, D. 2001, Rev. Mod. Phys., 73, 629
\bibitem[Lai \& Ho(2002)]{laiho02a} Lai, D. \& Ho, W.C.G. 2002, ApJ, 566, 373
 (\laiho)
\bibitem[Lai \& Ho(2003)]{laiho02c} Lai, D. \& Ho, W.C.G. 2003, ApJ, 588, 962
\bibitem[Lai \& Salpeter(1995)]{laisalpeter95} Lai, D. \& Salpeter, E.E.
 1995, Phys. Rev. A, 52, 2611
\bibitem[Lai \& Salpeter(1997)]{laisalpeter97} Lai, D. \& Salpeter, E.E.
 1997, ApJ, 491, 270
\bibitem[Lloyd(2003)]{lloyd03} Lloyd, D. 2003, MNRAS, submitted
 (astro-ph/0303561)
\bibitem[Mereghetti \etal(2002a)]{mereghettietal02a} Mereghetti, S.,
 Chiarlone, L., Israel, G.L., \& Stella, L. 2002a, in Proc. 270 WE-Heraeus
 Seminar on Neutron Stars, Pulsars, and Supernova Remnants,
 eds. Becker, W., Lesch, H., \& Tr\"{u}mper, J., (MPE Rep. 278; Garching: MPI),
 p.29
\bibitem[Mereghetti \etal(2002b)]{mereghettietal02b} Mereghetti, S.,
 De Luca, A., Caraveo, P.A., Becker, W., Mignani, R., \& Bignami, G.F. 2002b,
 ApJ, 581, 1280
\bibitem[M\'{e}sz\'{a}ros(1992)]{meszaros92} M\'{e}sz\'{a}ros, P. 1992,
 High-Energy Radiation from Magnetized Neutron Stars (Chicago:
 University of Chicago Press)
\bibitem[M\'{e}sz\'{a}ros \& Ventura(1979)]{meszarosventura79}
 M\'{e}sz\'{a}ros, P. \& Ventura, J. 1979, Phys. Rev. D, 19, 3565
\bibitem[Mori \& Hailey(2003)]{morihailey03} Mori, K. \& Hailey, C.J. 2003,
 ApJ, submitted (astro-ph/0301161)
\bibitem[Morii \etal(2003)]{moriietal03} Morii, M., Sato, R., Kataoka, J.,
 \& Kawai, N. 2003, PASJ, 55, L45
\bibitem[Motch\etal(2003)]{motchetal03} Motch, C., Zavlin, V.E., \&
 Haberl, F. 2003, A\&A, 408, 323
\bibitem[Patel \etal(2001)]{pateletal01} Patel, S.K., \etal\ 2001, ApJL,
 563, L45
\bibitem[Patel \etal(2003)]{pateletal03} Patel, S.K., \etal\ 2003, ApJ, 587, 367
\bibitem[Pavlov \& Gnedin(1984)]{pavlovgnedin84} Pavlov, G.G. \& Gnedin, Yu.N.
 1984, Sov. Sci. Rev. E, 3, 197
\bibitem[Pavlov \& Potekhin(1995)]{pavlovpotekhin95} Pavlov, G.G. \&
 Potekhin, A.Y. 1995, ApJ, 450, 883
\bibitem[Pavlov \etal(2002)]{pavlovetal02} Pavlov, G.G., Zavlin, V.E.,
 \& Sanwal, D. 2002, in Proc. 270 WE-Heraeus Seminar on Neutron Stars, Pulsars,
 and Supernova Remnants, eds. Becker, W., Lesch, H., \& Tr\"{u}mper, J.,
 (MPE Rep. 278; Garching: MPI), p.273
\bibitem[Pavlov \etal(1995)]{pavlovetal95} Pavlov, G.G.,
 Shibanov, Yu.A., Zavlin, V.E., \& Meyer, R.D. 1995, in Lives of
 the Neutron Stars, eds. Alpar, M.A., Kizilo\u{g}lu, \"{U}., \& van Paradijs, J.
 (Boston: Kluwer), p.71
\bibitem[Potekhin(1994)]{potekhin94} Potekhin, A.Y. 1994, J. Phys. B, 27, 1073
\bibitem[Potekhin(1998)]{potekhin98} Potekhin, A.Y. 1998, J. Phys. B, 31, 49
\bibitem[Potekhin \& Chabrier(2003)]{potekhinchabrier03} Potekhin, A.Y. \&
 Chabrier, G. 2003, ApJ, 585, 955 (\potcha)
\bibitem[Potekhin \& Pavlov(1997)]{potekhinpavlov97} Potekhin, A.Y. \&
 Pavlov, G.G. 1997, ApJ, 483, 414
\bibitem[Potekhin \etal(1999)]{potekhinetal99} Potekhin, A.Y.,
 Chabrier, G., \& Shibanov, Yu.A. 1999, Phys. Rev. E, 60, 2193
\bibitem[Ruder \etal(1994)]{ruderetal94} Ruder, H., Wunner, G., Herold, H.,
 \& Geyer, F. 1994, Atoms in Strong Magnetic Fields (Berlin: Springer-Verlag)
\bibitem[Sanwal \etal(2002)]{sanwaletal02} Sanwal, D., Pavlov, G.G.,
 Zavlin, V.E., \& Teter, M.A. 2002, ApJL, 574, L61
\bibitem[Thompson(2001)]{thompson01} Thompson, C. 2001, in
 Neutron Star-Black Hole Connection, eds. Kouveliotou, C., Ventura, J.,
 \& van den Heuvel, E. (Dordrecht: Kluwer), p.369
\bibitem[Thompson \& Duncan(1996)]{thompsonduncan96} Thompson, C.
 \& Duncan, R.C. 1996, ApJ, 473, 322
\bibitem[Tiengo \etal(2002)]{tiengoetal02} Tiengo, A., Goehler, E.,
 Staubert, R., \& Mereghetti, S. 2002, A\&A, 383, 182
\bibitem[Treves \etal(2000)]{trevesetal00} Treves, A., Turolla, R.,
 Zane, S., \& Colpi, M. 2000, PASP, 112, 297
\bibitem[Tsai \& Erber(1975)]{tsaierber75} Tsai, W.Y. \& Erber, T. 1975,
 Phys. Rev. D, 12, 1132
\bibitem[Zane \etal(2001)]{zaneetal01} Zane, S., Turolla, R., Stella, L., \&
 Treves, A. 2001, ApJ, 560, 384
\bibitem[Zavlin \& Pavlov(2002)]{zavlinpavlov02} Zavlin, V.E. \& Pavlov, G.G.
 2002, in Proc. 270 WE-Heraeus Seminar on Neutron Stars, Pulsars, and
 Supernova Remnants, eds. Becker, W., Lesch, H., \& Tr\"{u}mper, J.,
 (MPE Rep. 278; Garching: MPI), p.263

\end{thebibliography}
\end{document}